\documentclass{IEEEtran}
\usepackage{cite}
\usepackage{amsmath,amssymb,amsfonts}
\usepackage{graphicx}
\usepackage{textcomp,nicefrac}

\def\BibTeX{{\rm B\kern-.05em{\sc i\kern-.025em b}\kern-.08em
T\kern-.1667em\lower.7ex\hbox{E}\kern-.125emX}}
\begin{document}
\title{Design and Characterization of Racetrack 3D-Trench Silicon Sensor Based on 8-Inch Process with Excellent Time Resolution}
\author{Huimin Ji, Manwen Liu, Kuo Ma, Chuan Liao, Yanwen Liu, Zheng Li, Zhihua Li, and Jun Luo
\thanks{
This paper is submitted on \today.
This work was supported in part by the National Natural Science Foundation of China under Grant 12375188 and in part by the National Key Research and Development Program of China under Grant 2023YFF0719600. (Corresponding author: Manwen Liu and Zhihua Li)
}
\thanks{Huimin Ji, Manwen Liu, Zhihua Li, and Jun Luo are with the Key Laboratory of Fabrication Technologies for Integrated Circuits, Chinese Academy of Sciences, Beijing 100029, also with the Institute of Microelectronics, Chinese Academy of Sciences, Beijing 100029, China, and also with the School of Integrated Circuits, University of Chinese Academy of Sciences, Beijing 100049, China (e-mail: liumanwen@ime.ac.cn, lizhihua@ime.ac.cn).}
\thanks{Kuo Ma and Yanwen Liu are with the Department of Modern Physics and State Key Laboratory of Particle Detection and Electronics, University of Science and Technology of China, Hefei 230026, China.
}
\thanks{Chuan Liao is with the International Center for Quantum-Field Measurement Systems for Studies of the Universe and Particles (QUP), High Energy Accelerator Research Organization (KEK), Tsukuba 305-0801, Japan.}
\thanks{Zheng Li is with the School of Integrated Circuits, Ludong University, Yantai 264025, China, and also with the Institute of Microelectronics, Chinese Academy of Sciences, Beijing 100029, China.
}}

\maketitle

\begin{abstract}
In the extreme environments of high-luminosity colliders, traditional planar silicon sensors suffer severe radiation-induced performance degradation and fail to satisfy the stringent demands of high-precision tracking and high-speed timing in particle physics. 3D silicon sensors enhance radiation hardness by shortening charge collection distance, yet conventional designs with columnar or square-cell trench electrodes exhibit non-uniform electric fields, including saddle points and low-field regions, which degrade charge collection efficiency and timing resolution. This work presents a novel racetrack 3D-trench silicon sensor with continuous racetrack electrodes surrounding a long central collection electrode, aiming to eliminate electric field inhomogeneities. For the first time, a 23 $\mu $m shallow-etched device was fabricated on an 8-inch platform, which provides a promising basis for its subsequent mass production and engineering applications. The device performance was systematically evaluated through theoretical analysis, 3D TCAD simulations, and characterization using semiconductor parameter analyzers and transient current technique (TCT) measurements. The sensor achieves leakage current below 0.2 nA, breakdown voltage above 110 V, full depletion voltage as low as a few volts, capacitance as low as 650 fF, collected charge of 4 fC, time response of about 640 ps, and time resolution of 50 ps. This large-scale manufacturable, shallow-etched racetrack 3D-trench silicon sensor provides a competitive device solution for portable radiation detection and next-generation 4D tracking under high-radiation and high-event-rate conditions.
\end{abstract}


\section{Introduction}
\label{sec:introduction}
\IEEEPARstart{S}{ilicon} sensors are indispensable components in high-energy and nuclear physics experiments, renowned for their excellent spatial resolution and rapid response. 
However, the upcoming high-luminosity upgrades of the Large Hadron Collide (HL-LHC) will expose these sensors to significantly harsher radiation environments. 
The resulting radiation damage manifests as increased leakage current, carrier trapping, and degradation in charge collection efficiency, ultimately limiting the performance and lifespan of conventional planar silicon sensors.

To address these challenges, Parker \textit{et al.} introduced the concept of 3D columnar electrode sensors in 1997~\cite{PARKER1997328, 785737}, which fundamentally shortened carrier drift distances and reduced full depletion voltages. 
3D columnar silicon sensors are widely adopted in upgrades of major experiments such as ATLAS~\cite{DAVIA2012321, 10.3389/fphy.2021.624668}, owing to their relatively mature fabrication technology. However, conventional columnar electrodes with alternating n-type and p-type columns inevitably form potential saddle points at cell centers or boundaries due to geometric symmetry~\cite{Heijhoff:2023zvl}. In these regions, the electric field strength is severely depressed, forming low-field regions. The weak electric field leads to slow carrier drift, prolonged charge collection, and signal tailing. This degrades time resolution, and reduces charge collection efficiency~\cite{s25030926}.

To improve the uniformity of the electric field, the 3D trench electrode sensor design was proposed in 2009~\cite{LI201190}, featuring vertical trenches that surround a central electrode. This configuration offers a more uniform electric field and better radiation hardness compared to columnar designs. When designed as pixel arrays, trench-electrode sensors typically employ square or rectangular unit cells. However, recent studies~\cite{mi11070674, 11271512} have shown that such square-cell designs inevitably create new low-field regions at the cell corners due to their sharp right-angle geometry. The electric field strength in these corner regions is significantly lower than at the cell center, which again introduces non-uniform charge collection and degrades the detector’s timing performance. As radiation fluence continues to increase, further reduction in electrode spacing becomes essential, leading to constraints on cell size and sensor geometry. 

Against this backdrop, Liao \textit{et al.} proposed a novel trench electrode architecture that combines a cylindrical unit with a parallel plate unit in 2018~\cite{Liao2018/02, 10.1063/1.5042018}. Full 3D simulations demonstrated that this design can enlarge the sensor unit cell without introducing low electric field regions. 
Motivated by their work and constrained by available fabrication capabilities, we designed and successfully fabricated a novel racetrack 3D shallow trench silicon sensor. 
In contrast to Liao’s design, the sensor proposed in this work is fabricated on a thin p-type epitaxial layer instead of a bulk silicon substrate. This configuration effectively improves compatibility with standard CMOS process. Despite similarities in the overall device structure, our design exhibits substantial differences in key aspects, including the type and doping of charge collection electrode, trench geometric dimensions, and electrode spacing.

The key innovation of the proposed racetrack geometry lies in replacing the sharp right-angle corners of conventional square cells with smooth, continuous curved transitions. This design is intended to fundamentally eliminate low-field regions at corners and suppress the formation of potential saddle points by breaking the high geometric symmetry inherent in standard electrode arrays. As a result, a highly uniform electric field can be established across the entire active region, enabling fast, efficient, and spatially uniform charge collection. In this work, we elaborate on the design concept of the racetrack 3D-trench sensor and present corresponding TCAD simulation results. We then describe the device fabrication process and characterize its key performance, including dark current, capacitance, collected charge, time response, and time resolution. Finally, through preliminary evaluation of collected charge and time resolution, we demonstrate that the proposed racetrack 3D-trench sensor achieves great properties, offering a highly competitive solution to address the challenges of precision 4D tracking in future high-radiation environments.

\section{Design and Simulation of the Racetrack 3D-Trench Sensor}
\subsection{Design Theory}
\begin{figure}[t]
\centerline{\includegraphics[width=3.5in]{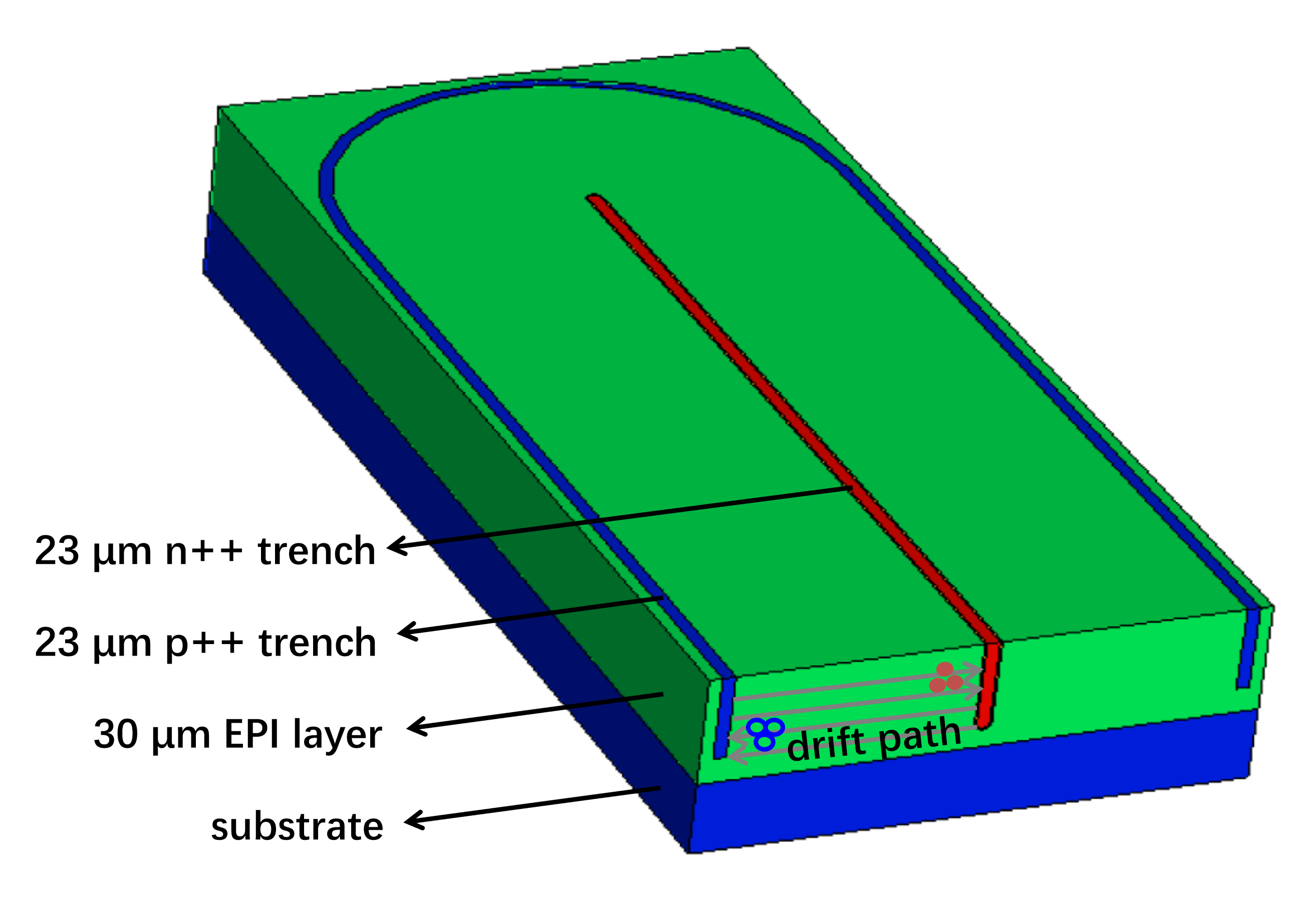}}
\caption{The schematic of the half Racetrack 3D-Trench Sensor.}
\label{structure}
\end{figure}
The core of the proposed device is a racetrack-shaped trench electrode, formed by two parallel straight segments connected by semicircular ends. \figurename~\ref{structure} presents a half view of the racetrack 3D-trench sensor structure to facilitate intuitive observation. The 30 $\mu $m thick p-type epitaxial layer is doped at approximately $1 \times 10^{12}$ cm$^{-3}$, which defines the effective doping concentration (N$_\text{eff}$). The substrate is p-type silicon with a doping concentration of $2.32 \times 10^{21}$ cm$^{-3}$. The trench is 3 $\mu $m wide and 23 $\mu $m deep. This depth allows the electrode to penetrate through most of the active region without fully breaking through the epitaxial layer. The central n$^{++}$ trench region is heavily doped with phosphorus at a concentration of $4.26 \times 10^{20}$ cm$^{-3}$, whereas the surrounding p$^{++}$ trench is heavily doped with boron at a concentration of $2 \times 10^{19}$ cm$^{-3}$. Both the p$^{++}$ and n$^{++}$ surfaces are coated with a 1 $\mu $m thick aluminum layer, whereas oxide layer covers the bulk surface to prevent short-circuit.

\begin{figure}[t]
\centerline{\includegraphics[width=3.5in]{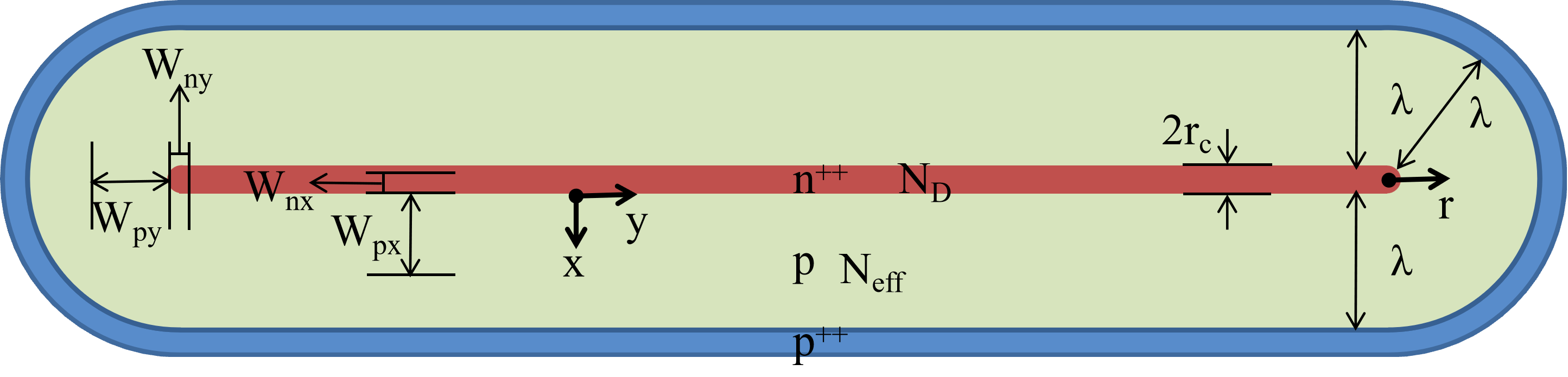}}
\caption{The cross-sectional diagram on the xy plane of the Racetrack 3D-Trench Sensor.}
\label{xycross}
\end{figure}

We could calculate the full depletion voltage of this device using the known parameters. \figurename~\ref{xycross} shows the cross-sectional diagram on the xy plane of the racetrack 3D-trench sensor. The width of the trench is 3 $\mu $m and the electrode spacing $\lambda$ is 57 $\mu $m. The length of the central n$^{++}$ electorde is 440 $\mu $m.
First, we solve the Poisson equation in the x-direction:
\begin{equation}
\label{poisson_x}
-\frac{\partial^2 \phi}{\partial x^2} = \frac{dE}{dx} = \frac{\rho(x)}{\varepsilon \varepsilon_0} =
\begin{cases}
\frac{e N_D}{\varepsilon \varepsilon_0}, & -W_{nx} \leq x <  0 \\
-\frac{e N_{\text{eff}}}{\varepsilon \varepsilon_0}, & 0 \leq x < W_{px}
\end{cases}
\end{equation}
where $N_D$ denotes the doping concentration of the center trench electrodes, $W_{nx}$ represents the depletion width of the n$^{++}$ parallel-plate electrode, $N_{\rm eff}$ is the effective doping concentration of the p-type epitaxial layer, and $W_{px}$ is the depletion width in the epitaxial layer.
When an additional reverse bias voltage is applied, the boundary conditions for the electric field and electric potential are:
\begin{equation}
\label{boundary_x}
\begin{cases}
E(-W_{nx})=E(W_{px})=0 \\
\phi (-W_{nx})=V_{\rm bi}+V \\
\phi (W_{px})=0
\end{cases}
\end{equation}
where $V$ denotes the absolute value of the applied reverse bias voltage and $V_{\rm bi}$ is the built-in potential. 
The distributions of the electric field and electric potential along the x-direction can be solved by using (\ref{poisson_x}) and (\ref{boundary_x}):
\begin{equation}
\label{electric_x}
E(x) =
\begin{cases}
0, & x \leq -W_{nx} \\
\frac{e N_D}{\varepsilon \varepsilon_0}(x+W_{nx}), & -W_{nx} \leq x <  0  \\
-\frac{e N_{\text{eff}}}{\varepsilon \varepsilon_0}(x-W_{px}), & 0 \leq x < W_{px} \\
0, & x \geq W_{px}
\end{cases}
\end{equation}

\begin{equation}
\label{potential_x}
\phi(x) =
\begin{cases}
V_{\rm bi} + V, & x \leq -W_{nx} \\
-\frac{e N_D}{2\varepsilon \varepsilon_0}(x+W_{nx})^2 + (V_{\rm bi} + V), & -W_{nx} \leq x <  0  \\
\frac{e N_{\text{eff}}}{2\varepsilon \varepsilon_0}(x-W_{px})^2, & 0 \leq x < W_{px} \\
0, & x \geq W_{px}
\end{cases}
\end{equation}

Considering the relationship N$_\text{eff}$ $\ll$ N$_D$, we can infer that W$_{nx}$ $\ll$ W$_{px}$, and W$_{nx}$ can be approximated to zero. When the epitaxial layer achieves full depletion, the electrode spacing $\lambda$ = W$_{px}$. By utilizing the continuity conditions of the electric field E(x) and the electric potential $\phi$(x) at all boundaries, we can calculate the dependence of the full depletion voltage V$_\text{fd}$ on the electrode spacing $\lambda$. When x=0:
\begin{equation}
\label{potential_x=0}
\phi(0) =-\frac{e N_D}{2\varepsilon \varepsilon_0}(W_{nx})^2 + (V_{\rm bi} + V_{fd}) = \frac{e N_{\text{eff}}}{2\varepsilon \varepsilon_0}(-W_{px})^2
\end{equation}

So we can obtain that:
\begin{equation}
\label{Vfd}
V_{fd} \cong \frac{e N_{\text{eff}}}{2\varepsilon \varepsilon_0}\lambda^2 - V_{\rm bi}
\end{equation}

Next, we solve the Poisson equation of the cylinder cell with cylindrical coordinate
(ignored $\theta$):
\begin{equation}
\label{poisson_ry}
\frac{1}{r} \frac{d[ r E(r)]}{dr} = \frac{\rho(r)}{\varepsilon \varepsilon_0} =
\begin{cases}
\frac{e N_D}{\varepsilon \varepsilon_0}, & r_c-W_{ny} \leq r <  r_c \\
-\frac{e N_{\text{eff}}}{\varepsilon \varepsilon_0}, & r_c \leq r < r_c+W_{py}\\
\end{cases}
\end{equation}
where $W_{ny}$ represents the depletion width of the n$^{++}$ cylindrical electrode, $W_{py}$ is the depletion width in the epitaxial layer, and $r_c$ denotes the radius of the cylindrical electrode, which is the half of the trench width.
When an additional reverse bias voltage is applied, the boundary conditions for the electric field and electric potential are:
\begin{equation}
\label{boundary_ry}
\begin{cases}
E(r_c-W_{ny})=E(r_c+W_{py})=0 \\
\phi (r_c-W_{ny})=V_{\rm bi}+V \\
\phi (r_c+W_{py})=0
\end{cases}
\end{equation}

The distributions of the electric field (\ref{electric_ry}) and electric potential (\ref{potential_ry}) for the cylindrical electrode along the y-direction can be solved by using (\ref{poisson_ry}) and (\ref{boundary_ry}):
\begin{equation}
\label{electric_ry}
E(r)=
\begin{cases}
0, & r\leq r_c-W_{ny}\\
\frac{eN_D}{2\varepsilon\varepsilon_0}\left[r-\frac{(r_c-W_{ny})^2}{r}\right],
& r_c-W_{ny} \leq r <  r_c\\
-\frac{eN_{\text{eff}}}{2\varepsilon\varepsilon_0}\left[r-\frac{(r_c+W_{py})^2}{r}\right],
& r_c \leq r < r_c+W_{py}\\
0,& r \geq r_c+W_{py}
\end{cases}
\end{equation}

\begin{figure*}
\begin{equation}
\label{potential_ry}
\phi(r)=
\begin{cases}
V_{\rm bi} + V, & r\leq r_c-W_{ny}\\
\frac{e N_D}{4\varepsilon\varepsilon_0}\left[(r_c-W_{ny})^2 - r^2- 2(r_c-W_{ny})^2\ln\frac{r_c-W_{ny}}{r}\right]+ (V_{\rm bi} + V),
& r_c-W_{ny} \leq r <  r_c\\
\frac{e N_{\rm eff}}{4\varepsilon\varepsilon_0}\left[r^2 - (r_c+W_{py})^2+2(r_c+W_{py})^2\ln\frac{r_c+W_{py}}{r}\right],
& r_c \leq r < r_c+W_{py}\\
0,& r \geq r_c+W_{py}
\end{cases}
\end{equation}
\end{figure*}

Given that N$_\text{eff}$ $\ll$ N$_D$, we deduce W$_{ny}$ $\ll$ W$_{py}$; hence, W$_{ny}$ can be approximated to zero. When the epitaxial layer becomes fully depleted, the electrode spacing $\lambda$ = W$_{py}$. By applying the continuity conditions for both the electric field E(r) and the electric potential $\phi$(r) at all boundaries, we obtain the dependence of the full depletion voltage V$_\text{fd}$ on the electrode spacing $\lambda$. At r=r$_c$, the potential is given by (\ref{potential_r=r$_c$}). Consequently, V$_\text{fd}$ can be calculated as shown in (\ref{Vfd_r}).
\begin{figure*}
\begin{align}
\label{potential_r=r$_c$}
\phi(r_c) &=\frac{e N_D}{4\varepsilon\varepsilon_0}\left[(r_c-W_{ny})^2 - r_c^2- 2(r_c-W_{ny})^2\ln\frac{r_c-W_{ny}}{r_c}\right]+ (V_{\rm bi} + V_{fd}) \notag\\
&=\frac{e N_{\rm eff}}{4\varepsilon\varepsilon_0}\left[r_c^2 - (r_c+W_{py})^2+2(r_c+W_{py})^2\ln\frac{r_c+W_{py}}{r_c}\right]
\end{align}
\end{figure*}

\begin{figure*}
\begin{equation}
\label{Vfd_r}
V_{fd} \cong \frac{e N_{\rm eff}}{4\varepsilon\varepsilon_0}\left[r_c^2 - (r_c+\lambda)^2+
2(r_c+\lambda)^2\ln\frac{r_c+\lambda}{r_c}\right] - V_{\rm bi}
\end{equation}
\end{figure*}

Based on the various parameters of this sensor mentioned in the previous text, we calculate that the both full depletion voltage are less than 10 V.

\subsection{3D Simulation}
\begin{figure}[t]
\centerline{\includegraphics[width=3.5in]{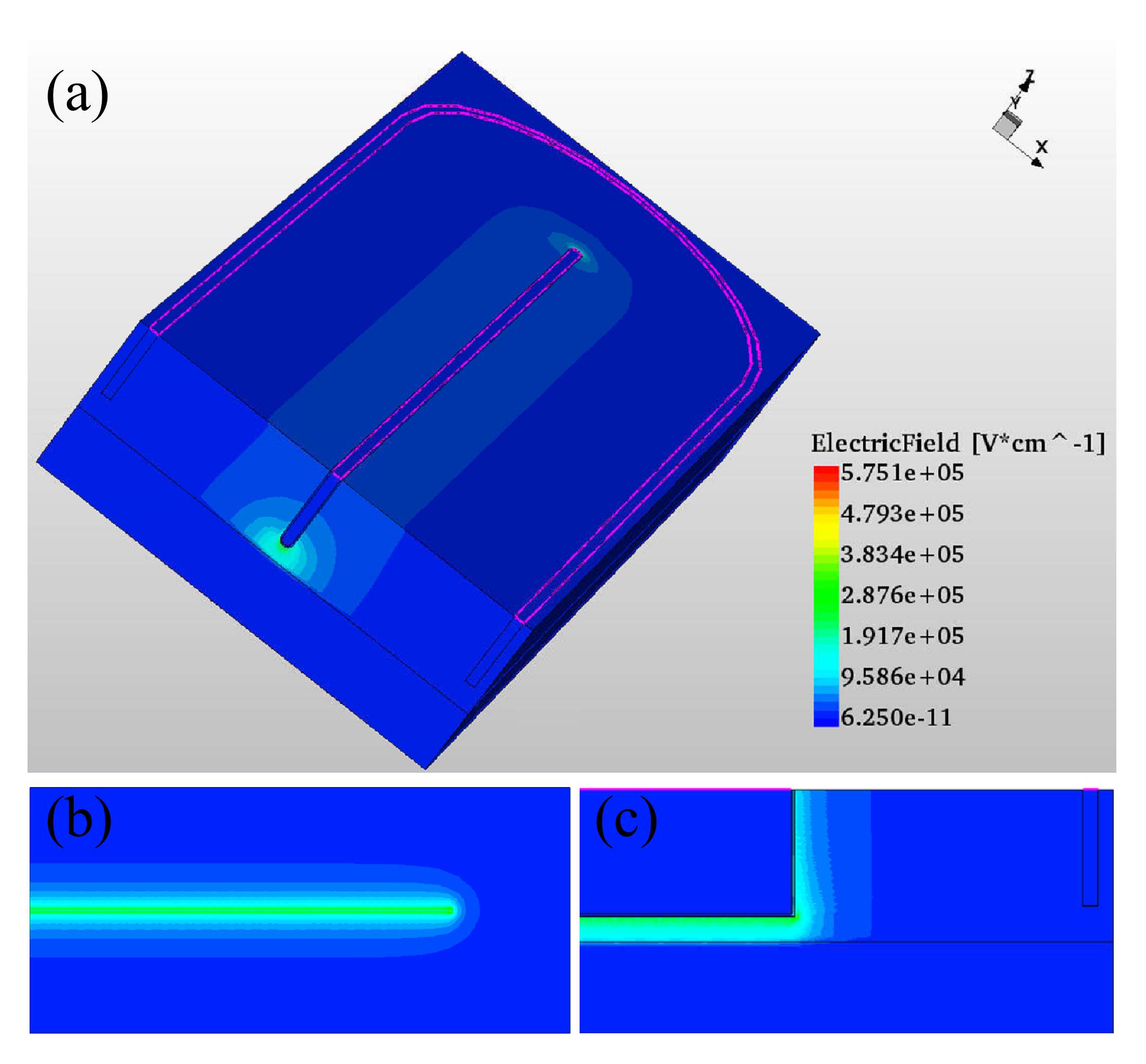}}
\caption{The simulated electric field in (a) 3D view, (b) XY plane at the depth of 25 $\mu $m, and (c) YZ plane at edge of the central electrode of the half Racetrack 3D-Trench Sensor at the reverse bias voltage of 90 V.}
\label{elec}
\end{figure}

\begin{figure}[t]
\centerline{\includegraphics[width=3.5in]{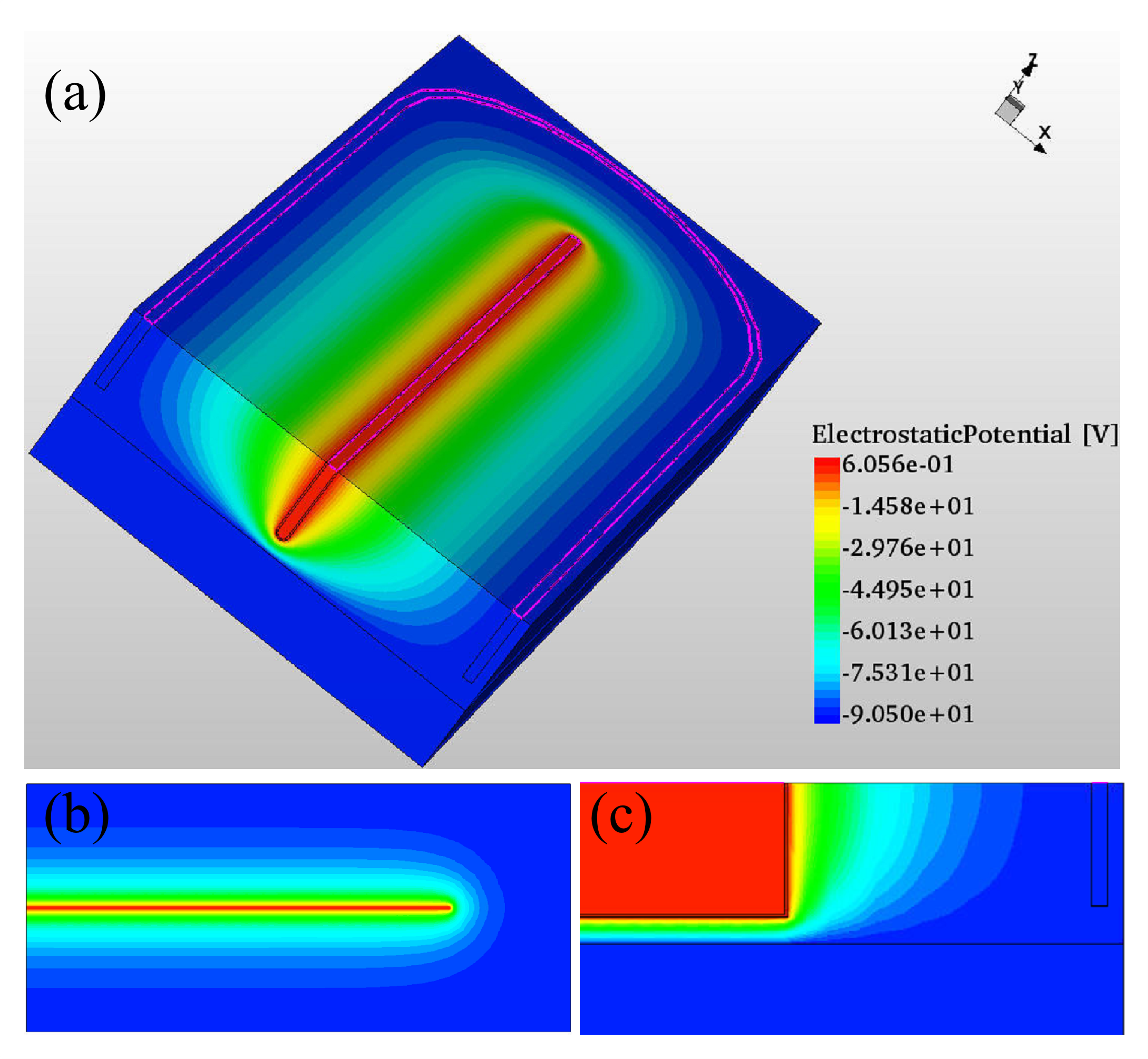}}
\caption{The simulated electrostatic potential in (a) 3D view, (b) XY plane at the depth of 25 $\mu $m, and (c) YZ plane at edge of the central electrode of the half Racetrack 3D-Trench Sensor at the reverse bias voltage of 90 V.}
\label{potential}
\end{figure}

\begin{figure}[t]
\centerline{\includegraphics[width=3.5in]{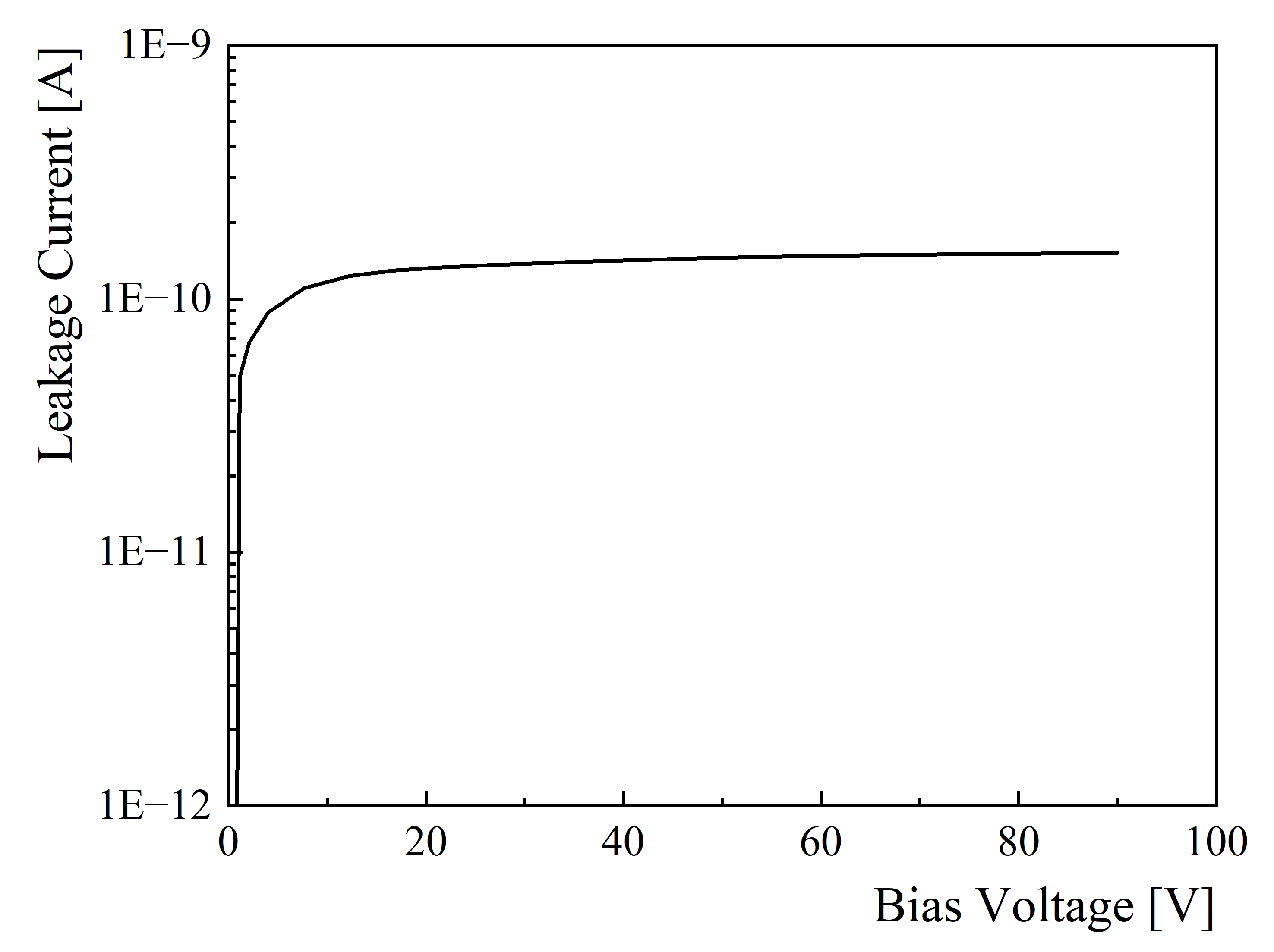}}
\caption{The simulated reverse I-V curves of the Racetrack 3D-Trench Sensor.}
\label{simIV90}
\end{figure}

\begin{figure}[t]
\centerline{\includegraphics[width=3.5in]{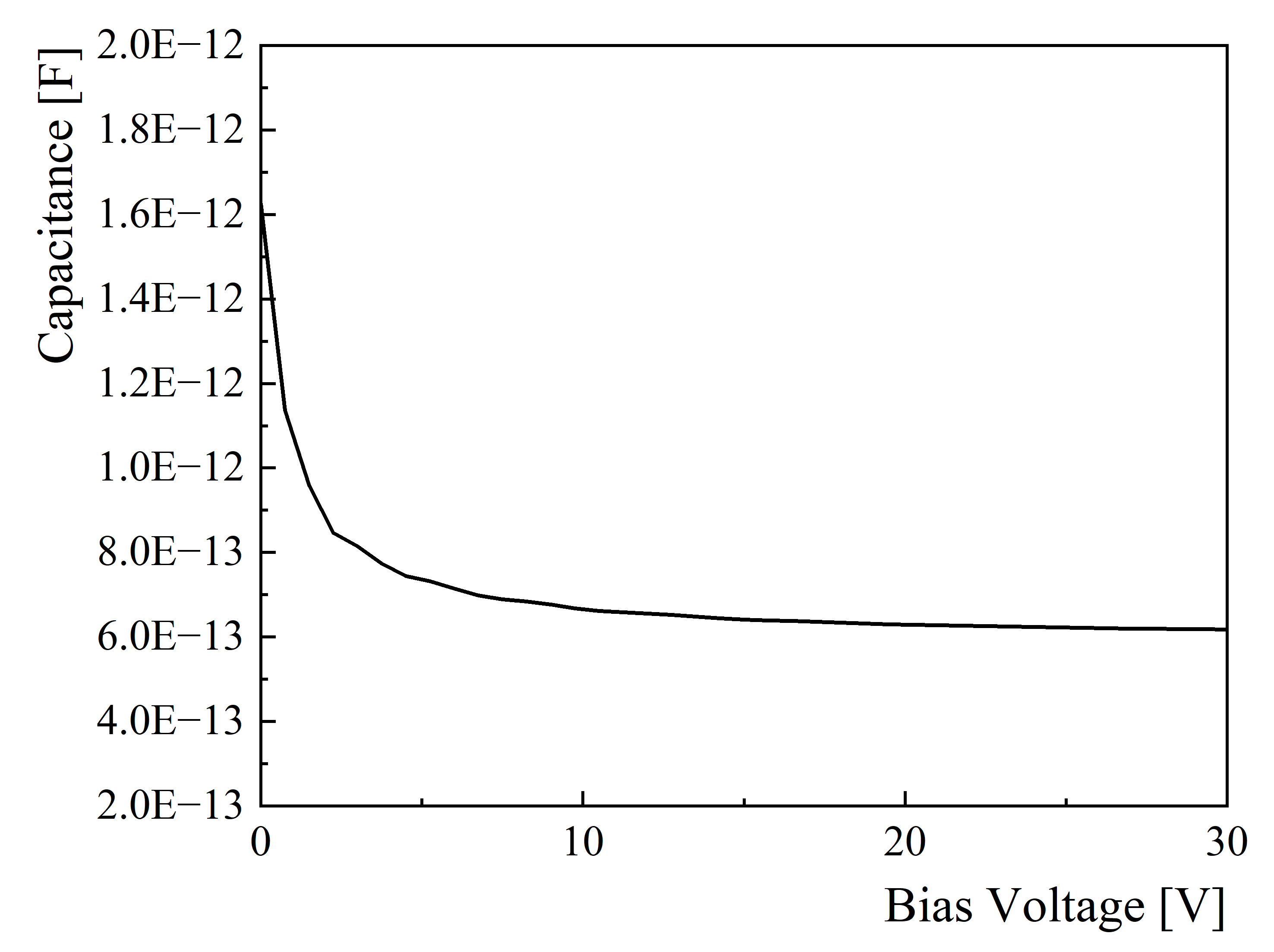}}
\caption{The simulated reverse C-V curves of the Racetrack 3D-Trench Sensor.}
\label{simCV30}
\end{figure}
We employed TCAD simulation tools to investigate the racetrack 3D-trench sensor, yielding the electric field distribution, potential distribution, leakage current, and capacitance characteristics. 
The interface charge density is set to 4$ \times 10^{11}$ cm$^{-2}$. Fermi statistics and the OldSlotboom effective intrinsic density are employed. Mobility models include doping dependence, high field saturation, and Enormal. Recombination‑generation is described by Shockley‑Read–Hall (SRH), Auger, and vanOverstraeten avalanche.  

\figurename~\ref{elec}a and \figurename~\ref{potential}a present the simulated distributions of the electric field and electrostatic potential in 3D view of the half racetrack 3D-trench sensor at the reverse bias voltage of 90 V.
\figurename~\ref{elec}b and \figurename~\ref{potential}b show the cutaway view of the simulated electric field and electrostatic potential in XY plane at the depth of 25 $\mu $m near the bottom of the central electrode with the reverse bias voltage of 90 V.
\figurename~\ref{elec}c and \figurename~\ref{potential}c show the cutaway view of the simulated electric field and electrostatic potential in YZ plane at edge of the central electrode with the reverse bias voltage of 90 V.
The TCAD simulation results strongly validate the correctness of our design concept. By eliminating sharp right-angle corners, the racetrack geometry successfully overcomes the issues of potential saddle points and corner low-field regions that exist in conventional 3D sensor designs. Its electric field uniformity lays a solid physical foundation for achieving fast, efficient, and uniform charge collection.

\figurename~\ref{simIV90} and \figurename~\ref{simCV30} shows the simulated leakage current and capacitance of the sensor. We observed that the simulated leakage current is about 0.1 nA, and the simulated capacitance is about 600 fF.

\section{Fabrication of the Racetrack 3D-Trench Sensor}
\begin{figure}[t]
\centerline{\includegraphics[width=3.5in]{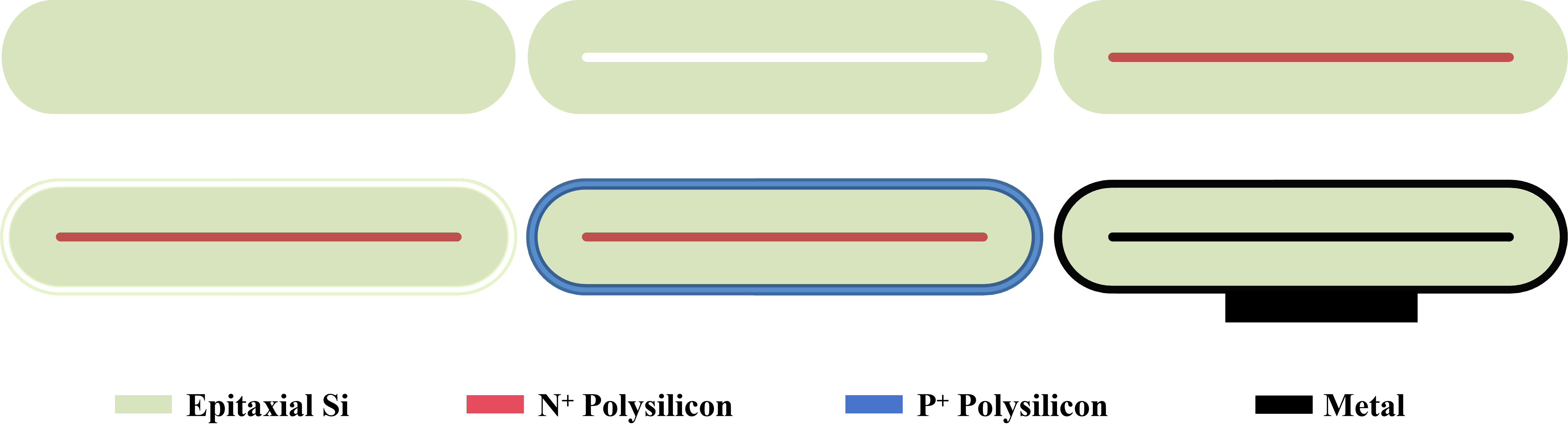}}
\caption{The simplified fabrication process flow of the Racetrack 3D-Trench Sensor.}
\label{fabrication}
\end{figure}

\begin{figure}[t]
\centerline{\includegraphics[width=3.5in]{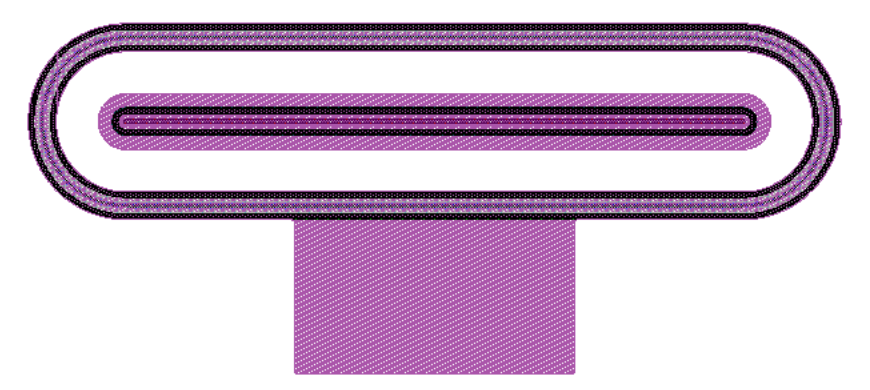}}
\caption{The layout of the Racetrack 3D-Trench Sensor.}
\label{layout}
\end{figure}

The racetrack 3D-trench sensors were fabricated on 8-inch epitaxial wafer based on the CMOS technology at Institute of Microelectronics of Chinese Academy of Sciences (IMECAS) in Beijing. The simplified fabrication process flow is shown in \figurename~\ref{fabrication}. First, we prepared a high-resistance epitaxial silicon wafer with an epitaxial thickness of 30 $\mu $m.
Next, we fabricated the oxide layer mask, spin-coated the photoresist, performed photolithography and development to create a 3 $\mu $m wide central long trench. This oxide layer in combination with the photoresist, acts as a hard mask during the subsequent etching process. Then, we use Deep Reactive Ion Etching (DRIE) with the Bosch process to create the trench with a depth of 23 $\mu $m based on the size of the opening feature. After center trench etching, we used in-situ doping of n-type polysilicon via Low-Pressure Chemical Vapor Deposition (LPCVD) to fill the trench. Then, the polysilicon in the other areas of the surface was etched away. Similarly, at the periphery of each device, we used DRIE with the Bosch process to create surrounding racetrack trench with a width of  3 $\mu $m and a depth of 23 $\mu $m. Then, we use in-situ doping of p-type polysilicon to fill the surrounding racetrack trench. Finally, the fabrication is completed by making metal electrodes for contact. \figurename~\ref{layout} present the layout of the fabricated sensor. The peripheral trench is the racetrack electrode, and the central long trench is the collection electrode. This layout eliminates the low-field regions formed at the corners of the electrodes in traditional columnar sensors, and theoretically enables the rapid collection of the charge.

\begin{figure}[t]
\centerline{\includegraphics[width=3.5in]{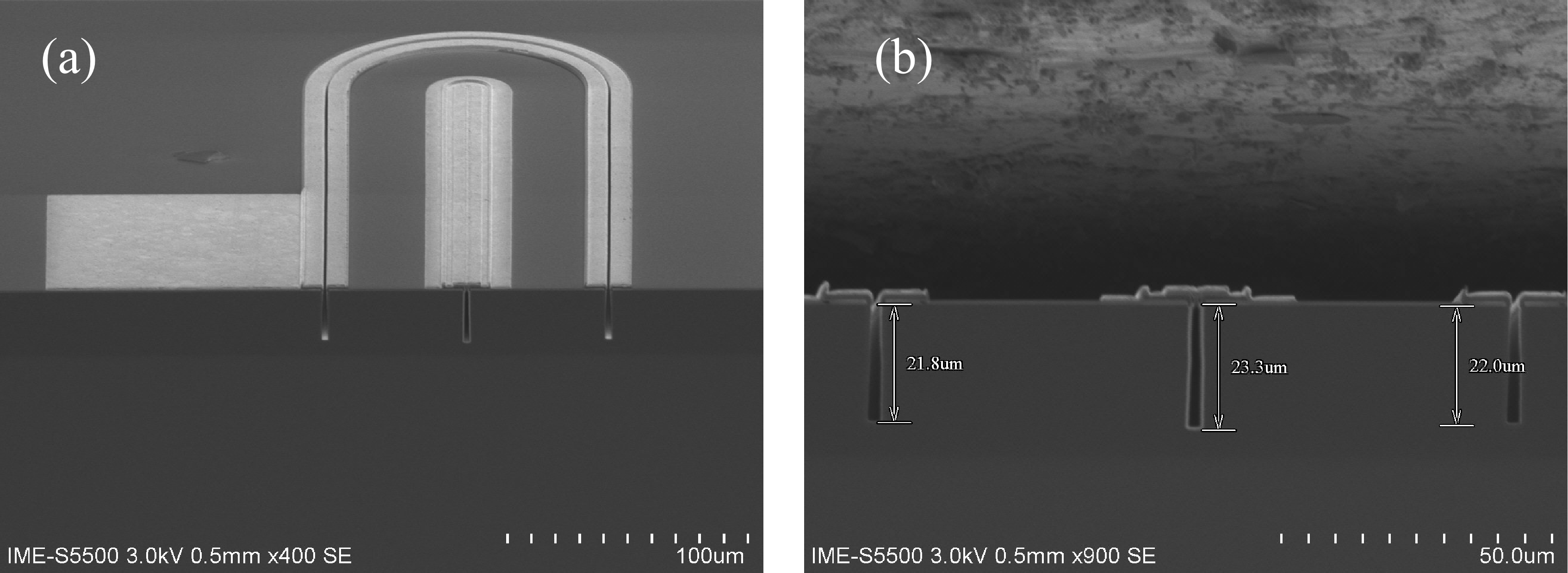}}
\caption{Surface (a) and section (b) views of the half Racetrack 3D-Trench Sensor.}
\label{sem}
\end{figure}

\figurename~\ref{sem}a shows the Scanning Electron Microscopy (SEM) micrography of the half racetrack 3D-trench sensor. \figurename~\ref{sem}b shows the morphology of the trenches. The etched trenches were not completely filled and the p$^{++}$ trench openings were not sealed. However, this does not affect the preparation of the device electrodes.

\section{Characterization Measurement of the Racetrack 3D-Trench Sensor}
\subsection{Electrical characteristics of the Device}
Current-voltage (IV) and capacitance-voltage (CV) characterizations were implemented in an ISO Class 6 cleanroom at the University of Science and Technology of China (USTC). The ambient temperature and relative humidity of the cleanroom were precisely controlled at 23 ± 1\,$^\circ$C and 38 ± 5\,\%, respectively. All electrical measurements were carried out utilizing an Apollowave alpha-200CS probe station.
Throughout the testing procedure, the sensors were placed on the vacuum chuck of the probe station, which immobilizes the sensor via a vacuum adsorption system. A reverse high voltage (HV) bias was applied to the p$^{++}$ trench electrode of the sensor through a KEITHLEY 2470 source measurement unit.
For the IV measurement, the n$^{++}$ trench electrode was connected to a KEITHLEY 6482 dual-channel picoammeter via probe needles for real-time current recording. For the CV measurement, the n$^{++}$ trench electrode was linked to an Agilent precision LCR meter (impedance analyzer, E4980A) via another probe needle, with an adapter installed in front of the LCR meter to isolate the high direct current voltage. 

\begin{figure}[t]
\centerline{\includegraphics[width=3.5in]{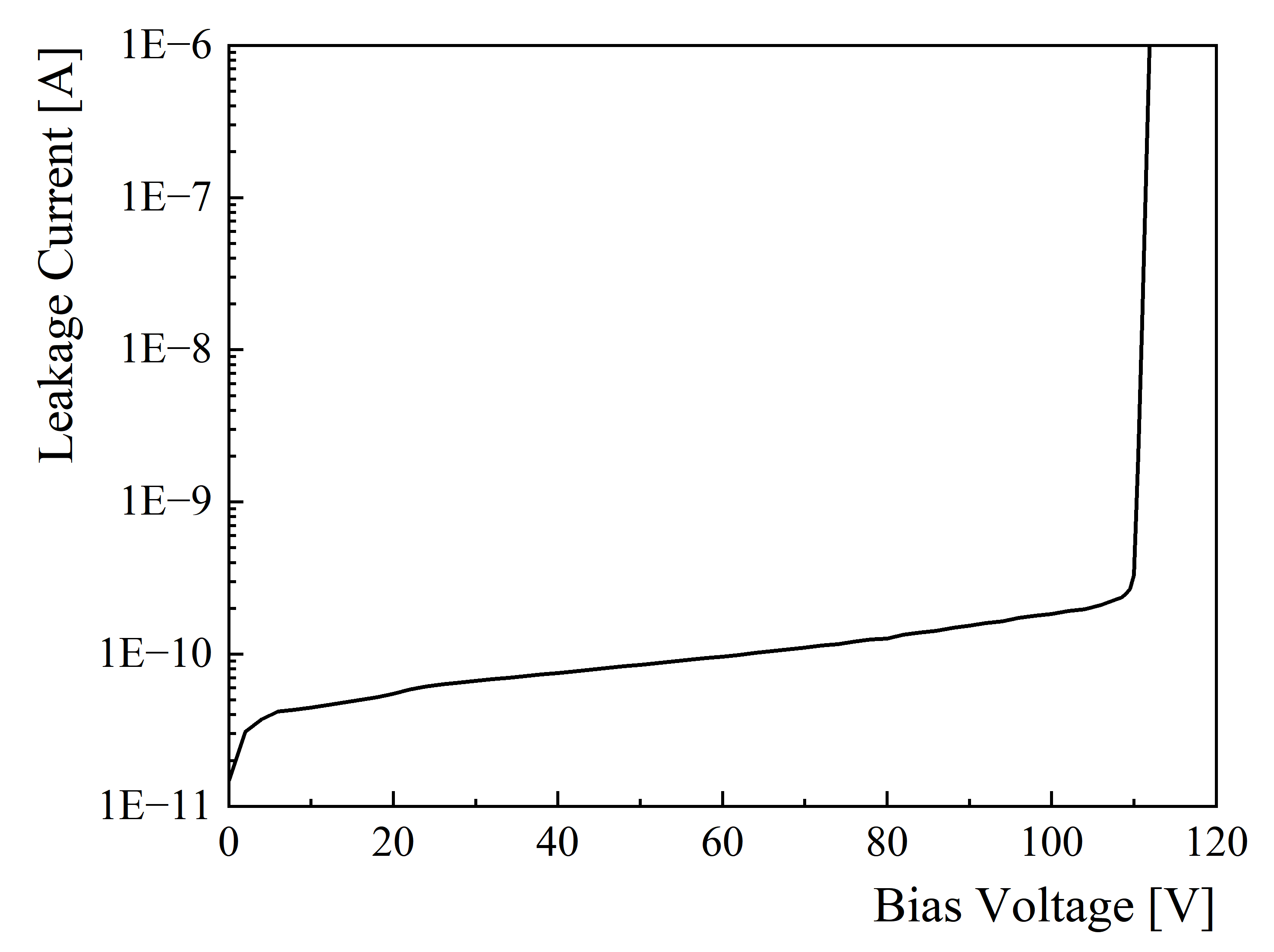}}
\caption{The leakage current versus reverse bias voltage (I-V) curve of the Racetrack 3D-Trench Sensor.}
\label{IVtest}
\end{figure}

\begin{figure}[t]
\centerline{\includegraphics[width=3.5in]{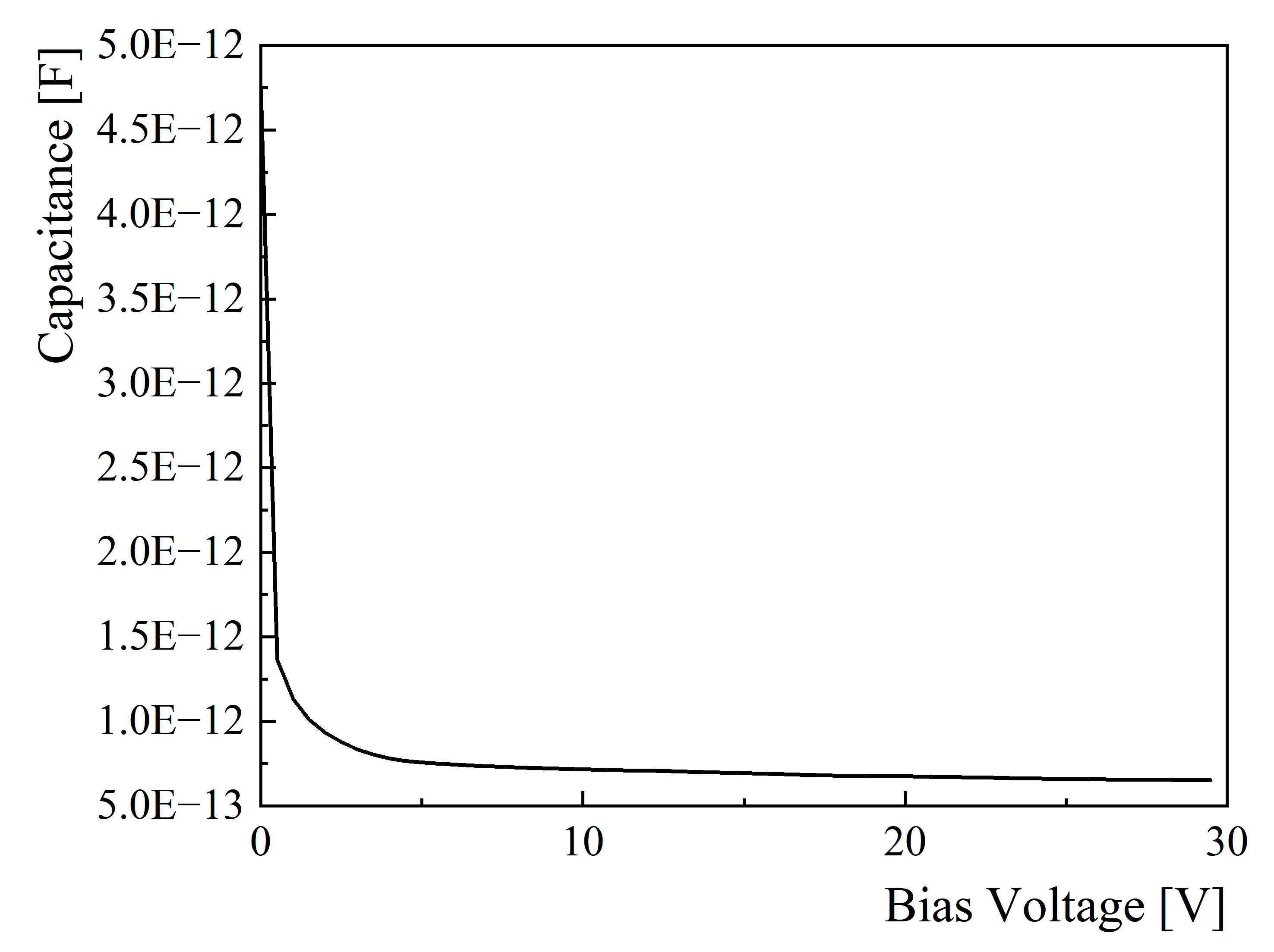}}
\caption{The capacitance versus revers bias voltage (C-V) curve of the Racetrack 3D-Trench Sensor.}
\label{CVtest}
\end{figure}

\figurename~\ref{IVtest} and \figurename~\ref{CVtest} show the measured IV and CV curves.
In the IV plot, we can observe that the leakage current of this device is less than 0.2 nA. Beyond a reverse bias of 110 V, the leakage current exhibits a pronounced rise, ultimately leading to the breakdown of the sensor.
In the CV plot, we can see the device has already been basically depleted at several volts. The capacitance is about 650 fF, which was conducted with an AC excitation signal of 51 mV amplitude at 10 kHz frequency. 

\subsection{Charge Collection and Time Resolution of the Device}
\begin{figure}[t]
\centerline{\includegraphics[width=3.5in]{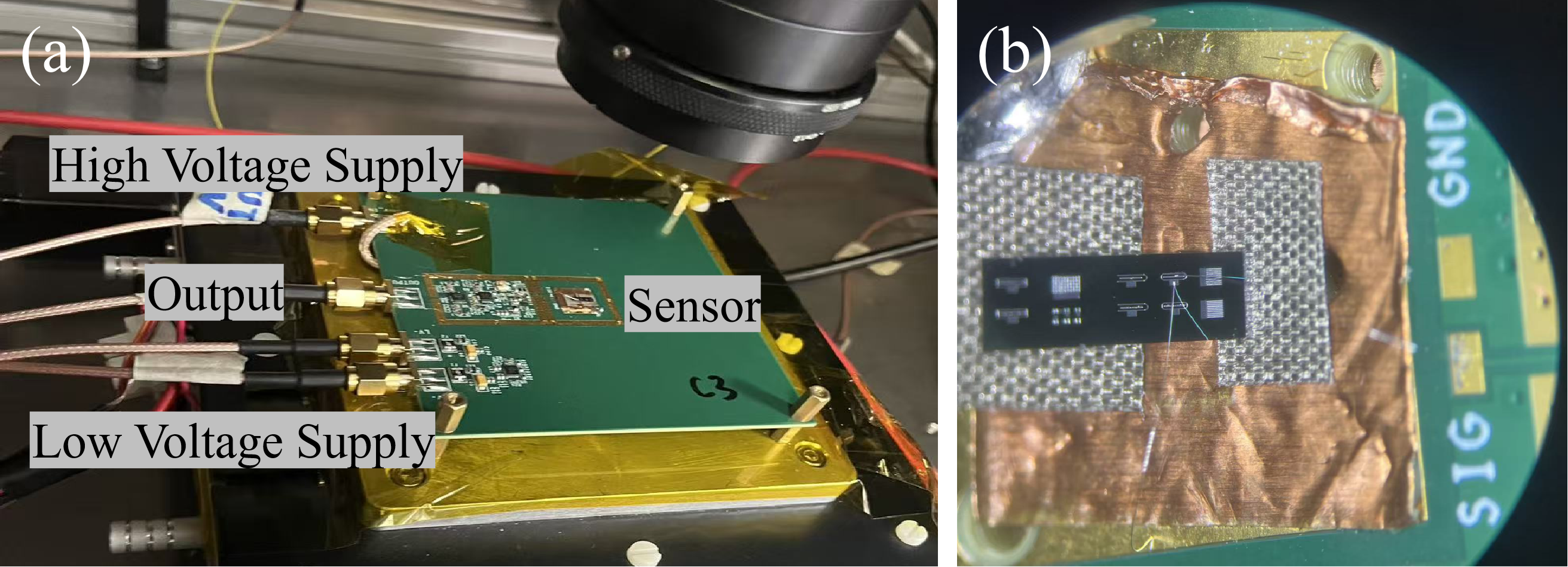}}
\caption{(a) The Racetrack 3D-Trench Sensor on the amplifier board in the TCT measurement system. (b) The microphotograph of the Racetrack 3D-Trench Sensor on the amplifier board.}
\label{TCT}
\end{figure}

To validate the performance of the proposed racetrack 3D-trench sensor, comprehensive characterizations were conducted at USTC utilizing infrared Transient Current Technique (TCT). The measurements were performed at room temperature using a customized infrared laser source operating at a wavelength of 1064 nm. The detailed introduction of the USTC amplifier board we used and the set-up of the TCT measurement system is presented in Ref.~\cite{1}.
As shown in \figurename~\ref{TCT}a, the amplifier board where the sensors are placed is positioned on the stage of the TCT measurement system. The laser spot, with a Full Width at Half Maximum (FWHM) of less than 11 $\mu $m, was focused perpendicularly onto the sensor surface, aligning it between the surrounding racetrack trench electrode and the central long trench electrode. Bias voltage was applied to the sensor via a Keithley 2470 high-voltage source meter, and the corresponding response signals were recorded using a Teledyne LeCroy HDO9404 high-definition oscilloscope. \figurename~\ref{TCT}b shows the connection between the sensor and the amplifier board under the microscope. The sensor was mounted onto the custom-designed amplifier board via double-sided conductive tape. Electrical connection was established through wire bonding on the signal pads.

\begin{figure}[t]
\centerline{\includegraphics[width=3.5in]{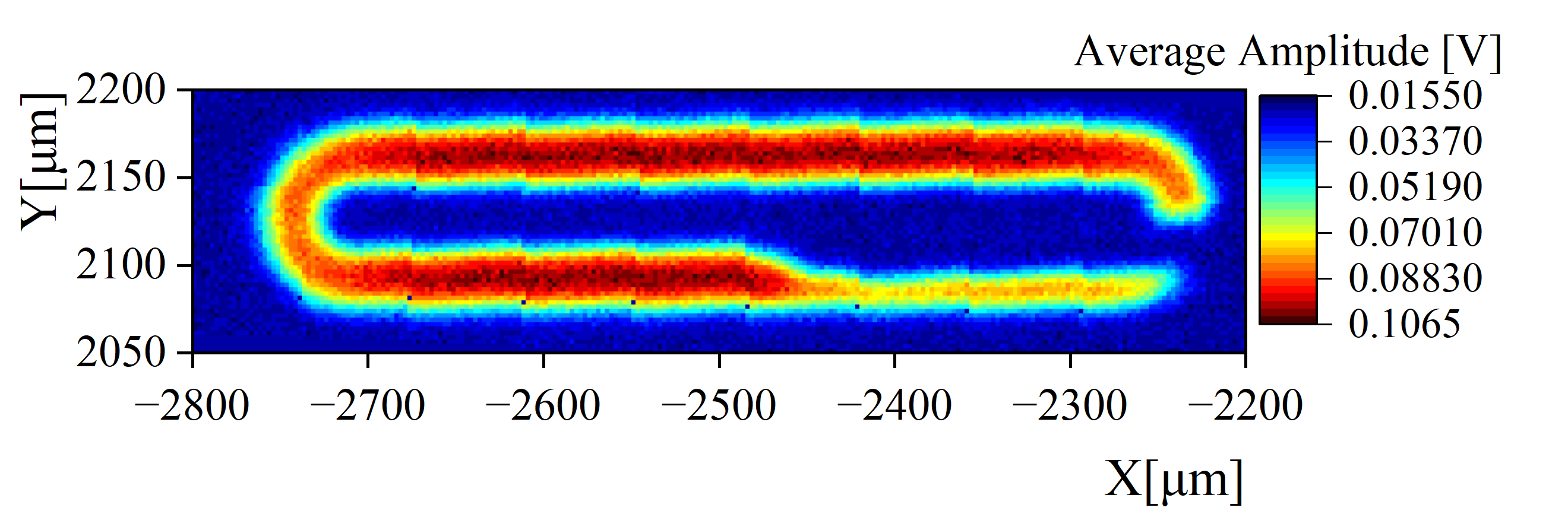}}
\caption{Maximum signal average amplitude of the Racetrack 3D-Trench Sensor with Laser70 injection at 80 V.}
\label{peak}
\end{figure}

\begin{figure}[t]
\centerline{\includegraphics[width=3.5in]{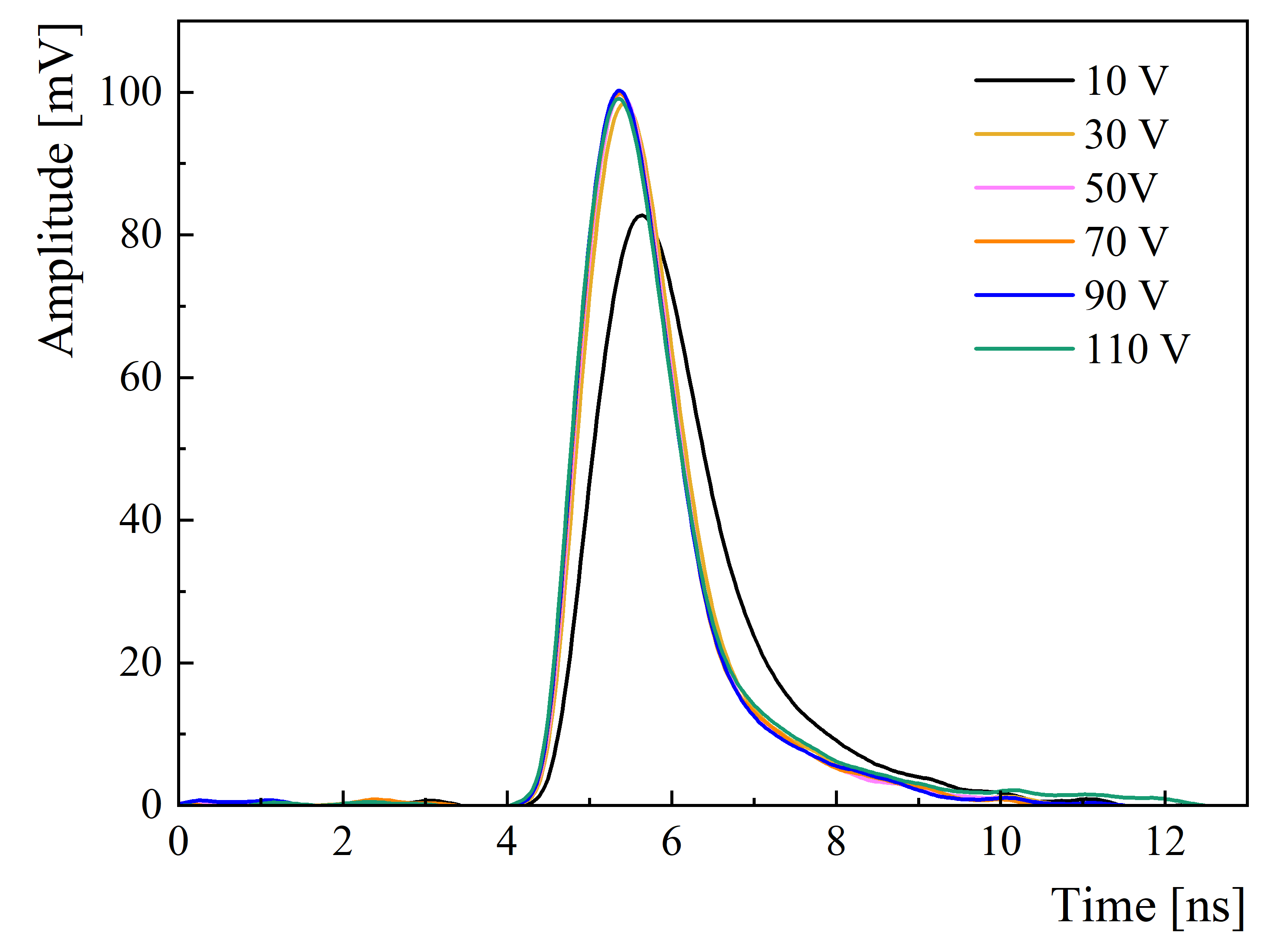}}
\caption{The average waveforms of the Racetrack 3D-Trench Sensor with Laser70 injection at different bias voltages.}
\label{ave}
\end{figure}

\begin{figure}[t]
\centerline{\includegraphics[width=3.5in]{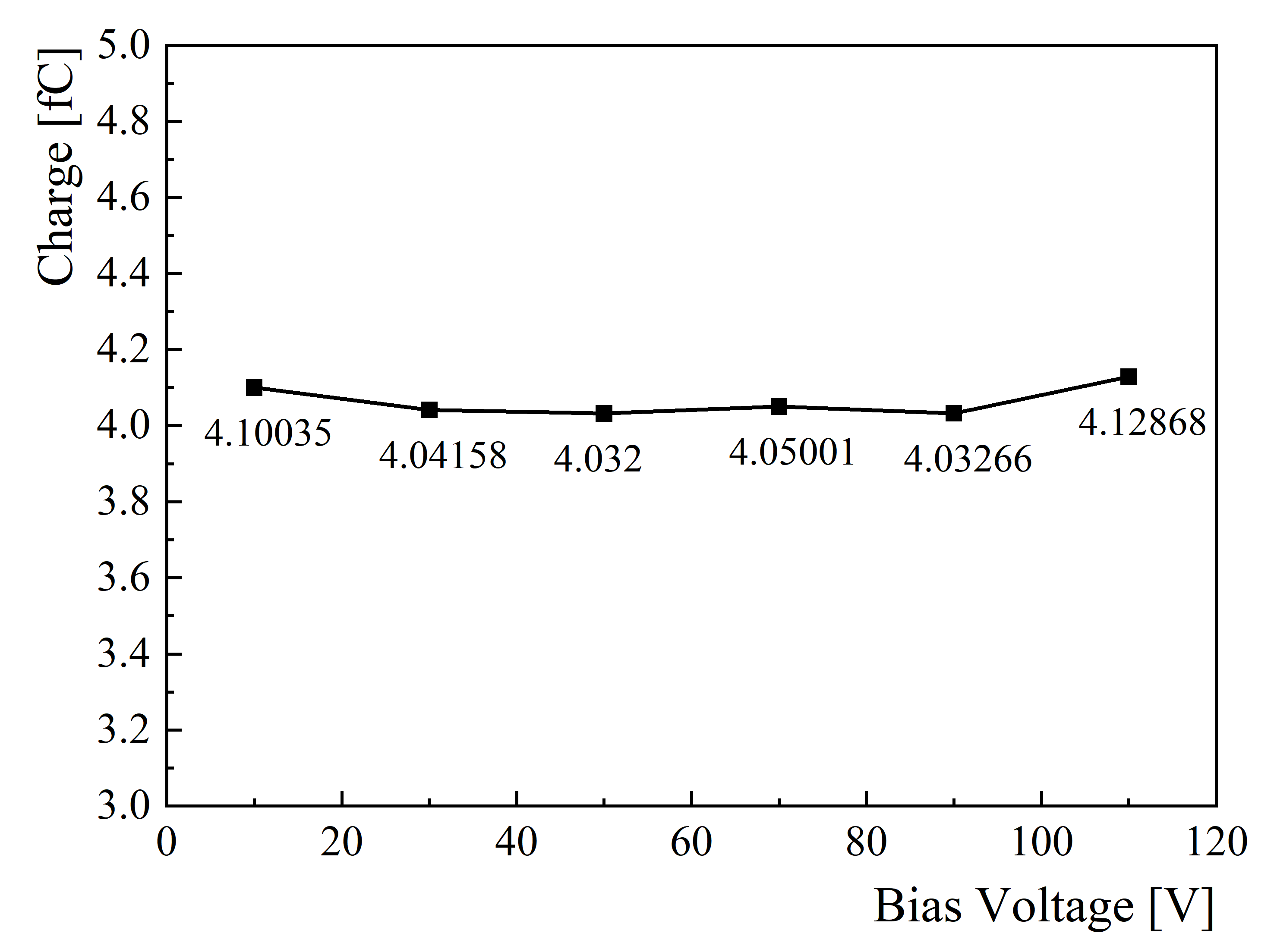}}
\caption{The collected charge of the Racetrack 3D-Trench Sensor with Laser70 injection at different bias voltages.}
\label{collectcharge}
\end{figure}

\begin{figure}[t]
\centerline{\includegraphics[width=3.5in]{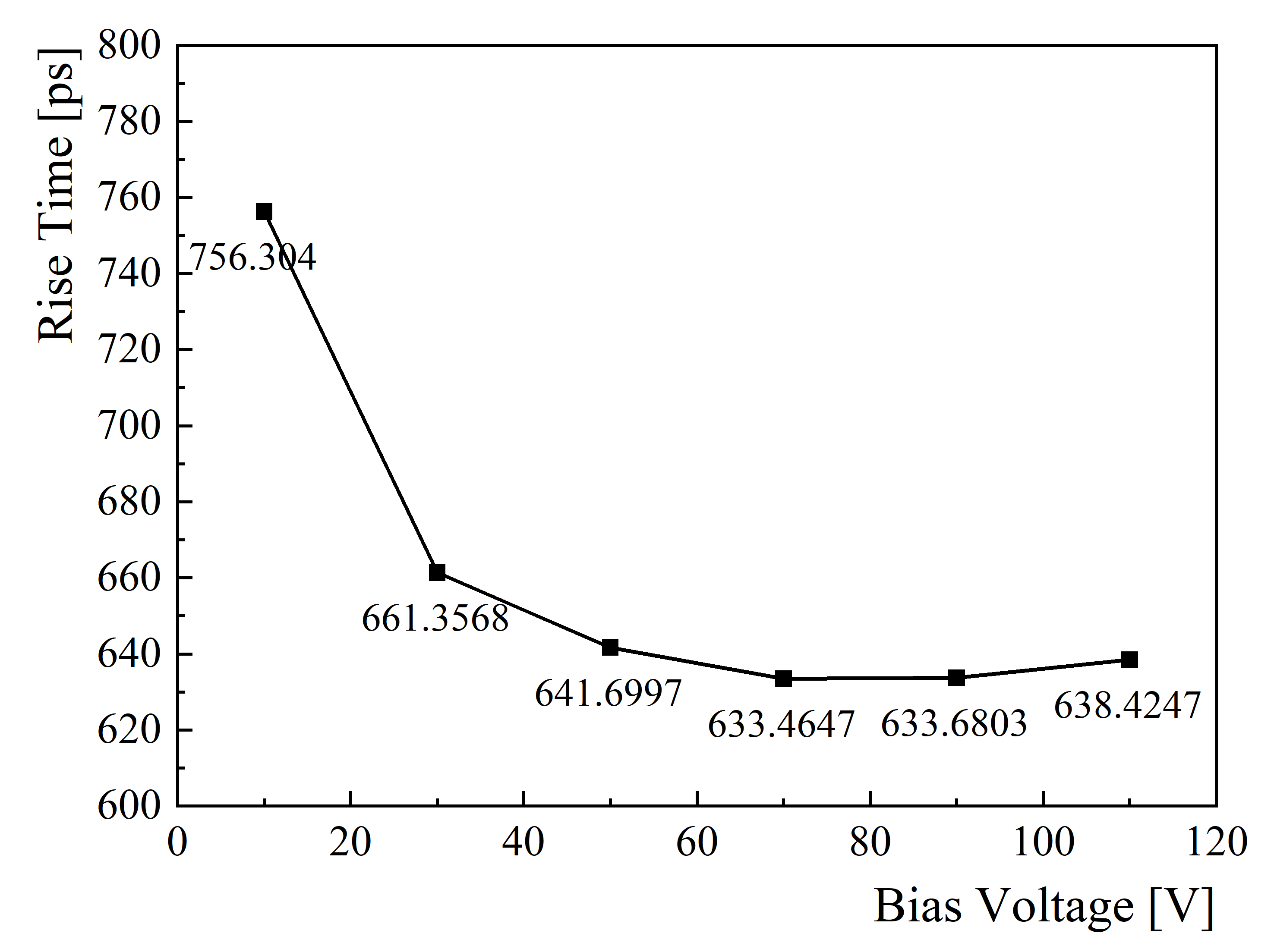}}
\caption{The rise time of the Racetrack 3D-Trench Sensor with Laser70 injection at different bias voltages.}
\label{risetime}
\end{figure}

\figurename~\ref{peak} shows the average peak amplitudes of the sensor at different positions with Laser70 injection when the reverse bias voltage is 80 V. As can be seen from the figure, the shape and size of the racetrack sensor are presented, which are close to the designed layout. Some areas are not shown because they are blocked by the wire between the central electrode and the signal pad.
Subsequently, we selected one of the regions with the higher amplitude for voltage scanning with Laser70 injection.
\figurename~\ref{ave} presents the average waveforms of the sensor under different bias voltages with Laser70 injection, which demonstrates that the sensor exhibits favorable functionality in charge collection measurements. After the reverse bias voltage exceeds 30 V, the average waveform remains stable.
The collected charge is determined by integrating the output pulses and dividing the integrated value by the calibration constant of the amplifier board. \figurename~\ref{collectcharge} shows the charge collection under different bias voltages. It can be observed that the collected charge does not increase as the reverse bias voltage is raised from 10 V to 110 V. The collected charge of the sensor is measured to be approximately 4 fC. This also indicates that the racetrack 3D-trench sensor is already fully depleted at approximately 10 V. 

\begin{figure}[t]
\centerline{\includegraphics[width=3.5in]{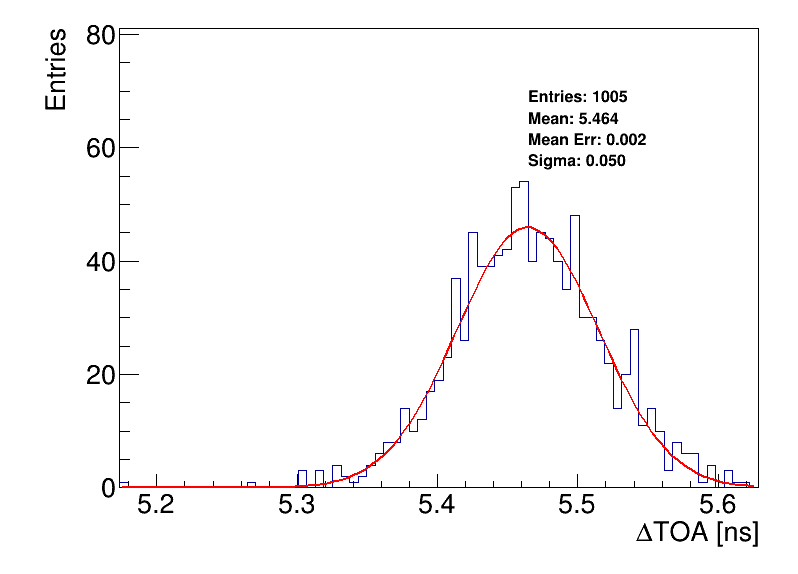}}
\caption{Distribution of the time-of-arrival difference ($\Delta$TOA) obtained from the racetrack 3D-trench sensor operated at 90 V. The y‑axis indicates the number of events per histogram bin (Entries), and the extracted sigma denotes the standard deviation of the $\Delta$TOA distribution.}
\label{toa_90}
\end{figure}

\begin{figure}[t]
\centerline{\includegraphics[width=3.5in]{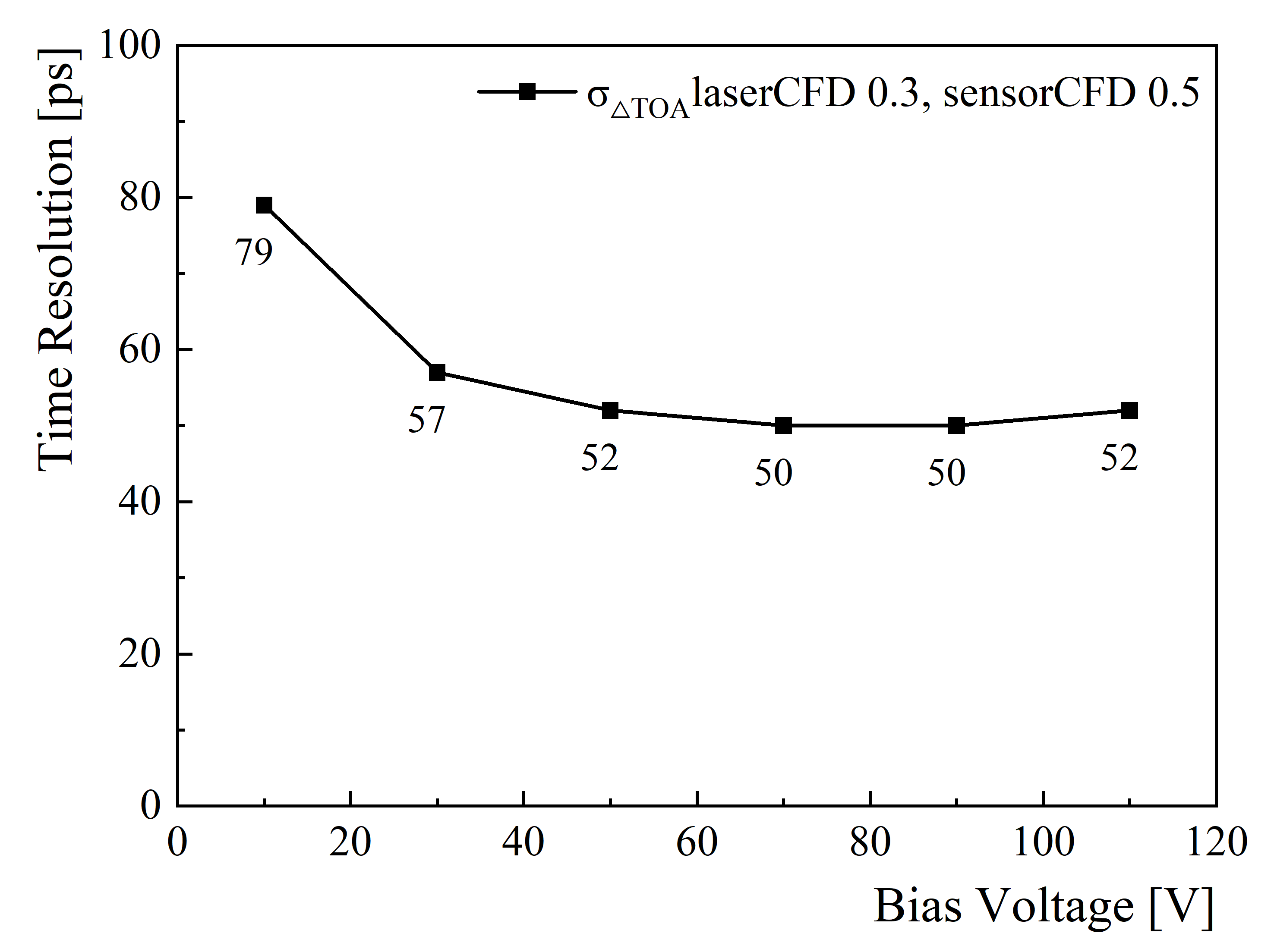}}
\caption{The time resolution of the Racetrack 3D-Trench Sensor with Laser70 injection at different bias voltages.}
\label{time}
\end{figure}

The time response refers to the rise time, which is equal to the time from 10\% to 90\% peak value. \figurename~\ref{risetime} displays the rise time of the sensor average waveforms with Laser70 injection at different bias voltages. At higher bias voltages, the rise time remains stable at around 640 ps. The reason why the rise time is higher at 10 V is mainly due to the low voltage, which results in insufficient electric field strength. 
The time-of-arrival (TOA) of the sensor signal and the laser synchronization signal were determined using the constant fraction discrimination (CFD) of around 30\% for the laser and 50\% for the sensor. This approach effectively suppresses the influence of time‑walk effect. Since there is only one laser pulse at the fixed position, the time resolution of the sensor can be approximately defined as the $\sigma$ value obtained after performing a Gaussian fit on the time‑of‑arrival difference ($\Delta$TOA) between the sensor and the laser pulse, which is independent of the time resolution of the laser system.
As shown in \figurename~\ref{toa_90}, the red line represents the Gaussian fit of $\Delta$TOA. The sigma value of the Gaussian fit of $\Delta$TOA at 90 V is 50 ps. \figurename~\ref{time} presents the time resolution of the racetrack 3D-trench sensor with Laser70 injection at different bias voltages. The time resolution after stabilization is approximately 50 ps.

\section{Conclusion}
A novel racetrack 3D-trench sensor was successfully designed and fabricated in this study. The compatible fabrication of the sensor was realized relying on the IMECAS 8-inch CMOS technology, and leakage current, capacitance, charge collection and time resolution testing were completed at USTC. The main conclusions are as follows:
\begin{itemize}
\item In terms of structural design, the  racetrack 3D-trench structure effectively addresses the inherent limitations of conventional sensors with columnar or square-cell trench electrodes, eliminating potential saddle points and low electric field regions at the device corners. Verified by TCAD simulation, the uniformity of electric field distribution is significantly better than that of traditional devices, providing a structural guarantee for great performance.
\item The device is fabricated using a 30 $\mu $m thick p-type epitaxial layer, achieving an etching width of 3 $\mu $m and etching depth of 23 $\mu $m. It is compatible with the 8-inch CMOS production process, with promising engineering application potential.
\item As for characterization measurements, the electrical test results show that the sensor has a dark current lower than 0.2 nA and a capacitance of about 650 fF, and its electrical characteristics match well with the simulation results. TCT measurements show that the collected charge is about 4 fC, the time response is about 640 ps, and the time resolution at the fixed measurement position can reach 50 ps, which is better than that of our previously designed 3D trench-column sensor~\cite{1, 2}.
\end{itemize}

The successful development of the racetrack 3D-trench silicon sensor demonstrates that optimizing electrode geometry to achieve a uniform electric field is an effective approach for enhancing the performance of 4D particle detectors. This design resolves the long-standing issue of electric field non-uniformity in 3D sensors and exhibits excellent timing performance, positioning it as a strong candidate for vertex and inner tracking detectors in the HL-LHC upgrade and future higher-energy colliders such as Future Circular Collider (FCC) and Circular Electron Positron Collider (CEPC).

Although this study has verified the functionality and core performance advantages of the racetrack 3D-trench sensor, there is still space for further optimization and in-depth research. we will focus on further optimizing the device structure and fabrication process, integrating the lessons accumulated in the production of this batch, improving the batch consistency and long-term stability of the device, and developing sensors with deeper trench. In future work, additional measurements will be implemented to fully characterize the device’s intrinsic performance, as well as investigate its long-term stability and radiation tolerance in diverse irradiation conditions. In addition, we will further explore its potential for ultra-fast timing and counting applications, expand its use in high-energy physics, astrophysics, and nuclear radiation detection, and advance the engineering and industrialization of racetrack 3D-trench radiation sensors.

\section*{Acknowledgment}
We are grateful for the support with equipment and technical personnel by the IMECAS and the USTC. In addition, this work performed in the framework of the CERN- DRD3 collaboration - WP2 project Novel silicon 3D-trench pixel detector fabricated on the 8-inch wafer utilizing CMOS processing technologies.

\bibliographystyle{IEEEtran}
\bibliography{reference}
\end{document}